\documentclass[aps,prd,eqsecnum,showpacs]{revtex4}
\newtheorem{tm}{Theorem}
\newtheorem{lm}{Lemma}
\newtheorem{dn}{Definition}
\usepackage{latexsym}
\usepackage{graphicx,subfigure}
\begin{document}
 
\thispagestyle{empty}
 
\title{Upper limits on the size of a primordial black hole}
\author{Tomohiro Harada\footnote{Electronic
address:T.Harada@qmul.ac.uk}
and 
B.~J. Carr\footnote{Electronic
address:B.J.Carr@qmul.ac.uk}}
\affiliation{
Astronomy Unit, School of Mathematical Sciences, 
Queen Mary, University of London, 
Mile End Road, London E1 4NS, UK}
\date{\today}
 
\begin{abstract}                
We provide precise constraints on the size of any black holes forming in the early Universe for a variety of formation scenarios. In particular, we prove that the size of the apparent horizon 
of a primordial black hole formed by 
causal processes in a flat Friedmann universe
is considerably smaller than the 
cosmological apparent horizon size
for an equation of state $p=k\rho$ ($1/3<k<1$).
This also applies for a stiff equation of state ($k=1$) 
or for a massless scalar field.
The apparent horizon of a primordial black hole  
formed through hydrodynamical processes
is also considerably smaller than the 
cosmological apparent horizon for $0<k\le 1$. We derive an expression
 for the maximum size which an overdense region can have without 
being a separate closed universe rather than part of our own.
Newtonian argument shows that a black hole smaller than the 
cosmological horizon can never accrete much.
\end{abstract}
\maketitle

\section{Introduction}
In the early stage of the universe,
black holes may have formed from
cosmological density perturbations~\cite{hawking1971}. 
Such ``primordial'' black holes (PBHs) may produce 
$\gamma$-ray background radiation and cosmic rays through quantum evaporation or
contribute to the density of the universe,
and this leads to observational constraints
on their number density~\cite{carr2003}. 
Studying the formation and evolution of PBHs
therefore leads to important constraints on models of the early universe.
However, to calculate these constraints, 
one needs to know how large PBHs were at their
formation epoch and how much they subsequently increased their mass through accretion.
 
A simple physical argument suggests that the size of a 
PBH will initially be comparable with the 
cosmological particle horizon scale for a hard equation 
of state~\cite{carr1975}. This is because an overdense region at maximum expansion must be larger than the Jeans 
scale in order to collapse against the pressure but not so large that it forms a separate closed universe.
Since both these scales are comparable to the particle horizon size, with the first being only slightly below the second, the size at maximum expansion must be finely tuned if a PBH is to form. (For a soft equation of state, the Jeans scale is 
much smaller than the cosmological horizon scale,
but the collapse is likely to lead to pancake formation and fragmentation unless the region is very spherically symmetric~\cite{khlopov1980}, so one again needs fine-tuning.) 
Since the size of a region at maximum expansion is determined by its initial overdensity (compared to the flat Friedmann background), only density fluctuations with a narrow range of amplitudes can produce PBHs. 

The limits on the size of a PBH have only been estimated rather roughly in previous papers. However, the fraction of the Universe going into PBHs is very sensitive to these limits~\cite{carr1975}, especially if the fluctuations on a given scale have a Gaussian distribution, so one would like to calculate them more precisely. In order to do this, 
we first need to determine the various cosmological scales that arise in
the early Universe: in particular, the cosmological particle horizon,
the cosmological apparent horizon, the cosmological sonic horizon and
the Jeans length. Depending on the formation scenario, these may all be
relevant to PBH production. They all have the same order of magnitude
but they also depend on the equation of state parameter $k$ (where
$p=k\rho$)
and one purpose of this paper is to calculate these dependencies for an exact Friedmann model very precisely.

The region undergoing collapse will not itself be part of a flat
Friedmann model, so in order to discuss the size of a PBH, one needs to
make some assumption about the form of the overdensity. One approach is
to assume that the universe is exactly Friedmann beyond some matching
radius. In this case, the black hole apparent horizon is necessarily
within the matching radius but its 
precise position (relative to some cosmological scale) depends on the density profile in the overdense region. The simplest models have constant overdensity, in which case the region can be regarded as part of a closed Friedmann model. However, more complicated possibilities can be envisaged, so it is important to obtain more general results. 

In this paper we will derive upper limits on the black hole size which are independent of any assumption about the density profile. In particular, we prove that the black hole apparent horizon
size is considerably smaller than the cosmological 
apparent horizon scale for an equation of state parameter in the range
$1/3<k\le 1$.
An important tool in proving these results is the use of double-null
coordinates. This is partly because, with the highly non-linear
fluctuations required for PBH formation, it is not always clear how to
construct spatial hypersurfaces of constant time. It is also because
double-null coordinates are convenient for numerical
calculations. Although we do not present such calculations here, we do
present them in an accompanying paper~\cite{hc2004}
for the case in which the universe contains a scalar field. 

If the fluctuations which generate the PBHs derive from inflation or
quantum gravity effects, they may effectively extend to infinity (or at
least well beyond the usual Friedmann particle horizon) rather than
being confined to a finite region.  In this case, the above results are
inapplicable and the only firm upper limit on the size of the PBH comes
from the separate universe condition. 
Although this constraint has been discussed in general 
terms~\cite{bm1992}, it cannot be formulated precisely without
specifying the density profile within the overdense region. However, we
will show that it can be calculated exactly in terms of the equation of
state parameter $k$ if the overdense region is assumed to be homogeneous.

The amount of accretion by a PBH is also sensitive to its initial size. A simple Newtonian calculation~\cite{zn1967} suggests that accretion is unimportant 
if the PBH is much smaller than the cosmological horizon at formation but that 
one could have self-similar
evolution, with the black hole growing at the same rate as the universe, if the size is initially comparable with 
the cosmological horizon scale.
The standard formation criterion implies that this is indeed the case, which suggests that there could be a large amount of accretion
for a hard equation of state.
However, self-similar PBH solutions have been
studied extensively~\cite{ch1974,lcf1976,bh1978a,bh1978b,cy1990} and it has been 
found that no such solution contains
a black hole inside an exact flat Friedmann solution 
for a perfect fluid with $0\le k\le 1$.
This applies even for a stiff fluid with $k=1$, although this has been the subject of some controversy~\cite{lcf1976,bh1978a}.
These results have been confirmed by numerical calculations~\cite{ nnp1978,np1980,nj1998,ss1999}. Indeed simulations 
suggest that, the harder the equation 
of state, the more difficult it will be 
for a PBH to form and the smaller the mass accretion rate will be.
Also one can show that a relativistic effect suppresses the accretion
when the PBH has a size very close to that of the cosmological 
apparent horizon~\cite{hc2004}.
\section{Double-Null Formulation of Einstein's Field Equations}
If we adopt units in which $G=c=1$ and the abstract index notation 
of~\cite{wald1983},
the Einstein equations
\begin{equation}
R_{ab}-\frac{1}{2}g_{ab}R=8\pi T_{ab}
\label{eq:einstein}
\end{equation}
can be written as
\begin{equation}
R_{ab}=8\pi \bar{T}_{ab},
\label{eq:einstein'}
\end{equation}
where 
\begin{eqnarray}
\bar{T}_{ab}&\equiv& T_{ab}-\frac{1}{2}g_{ab}T, \\
T&\equiv & T^{a}_{~a}.
\end{eqnarray}
We consider a spherically symmetric system, for which
the line element can be written in the form
\begin{equation}
ds^{2}=-a^{2}(u,v)dudv+r^{2}(u,v)(d\theta^{2}+\sin^{2}\theta d\phi^{2}),
\end{equation}
where $u$ and $v$ are advanced and retarded times, respectively. Then, from Eq.~(\ref{eq:einstein'}),
we obtain the following partial differential equations:
\begin{eqnarray}
\frac{r_{,uu}}{r}&=&2\frac{a_{,u}}{a}\frac{r_{,u}}{r}-4\pi \bar{T}_{uu}, 
\label{eq:Ruu}\\
\frac{r_{,uv}}{r}&=&\frac{a_{,u}a_{,v}}{a^{2}}-\frac{a_{,uv}}{a}
-4\pi \bar{T}_{uv}, 
\label{eq:Ruv}\\
\frac{r_{,vv}}{r}&=&2\frac{a_{,v}}{a}\frac{r_{,v}}{r}-4\pi 
\bar{T}_{vv}, 
\label{eq:Rvv}\\
\frac{r_{,uv}}{r}&=&-\frac{a^{2}}{4r^{2}}-\frac{r_{,v}r_{,u}}
{r^{2}}+\frac{a^{2}}{2r^{2}}
4\pi \bar{T}_{\theta\theta}.
\label{eq:Rtt}
\end{eqnarray}
There are no other independent components of the stress-energy tensor:
spherical symmetry requires 
$\bar{T}_{\theta\theta}=\sin^{2}\theta\bar{T}_{\phi\phi}$
and $\bar{T}_{\theta\phi}=0$. 
The initial data for the above equations are most naturally given 
on two initial null surfaces, $u=u_{0}$ and $v=v_{0}$. On each null surface the geometry is described by the two functions $a$ and
$r$.
 
The Hawking mass $m$, which is a well-behaved
quasi-local mass in spherically symmetric spacetimes, 
is defined as
\begin{equation}
m\equiv \frac{r}{2}\left(1+\frac{4r_{,u}r_{,v}}{a^{2}}\right).
\label{eq:misner-sharp}
\end{equation}
Equations~(\ref{eq:Ruu})--(\ref{eq:Rtt}) then imply 
(see also~\cite{hayward1996})
\begin{eqnarray}
m_{,u}&=&\frac{8\pi r^{2}}{a^{2}}( T_{uv}r_{,u}- T_{uu}r_{,v})
\label{eq:mu}\\
m_{,v}&=&\frac{8\pi r^{2}}{a^{2}}(T_{uv}r_{,v}- T_{vv}r_{,u})
\label{eq:mv}
\end{eqnarray}
We adopt the double null coordinate 
system as a natural choice of coordinates because
we will be dealing with highly nonlinear perturbations
in the cosmological background, so it is not always guaranteed that 
there exists a ``cosmologically preferred'' 
comoving time slicing. 
 
\section{Upper limit on black hole apparent horizon size}
\begin{figure}[htbp]
\begin{center}
\includegraphics{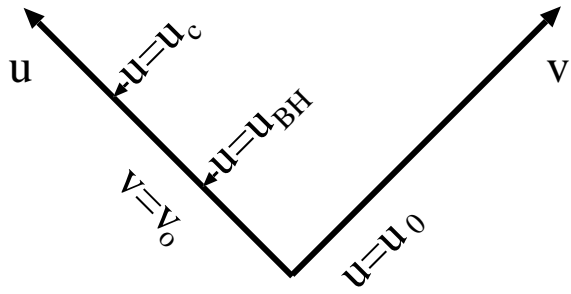}
\caption{\label{fg:null}
Schematic figure for proofs of 
Theorems~\ref{tm:r_decreasing}--\ref{tm:upper_limit_bh} 
and Lemma~\ref{lm:bhah}. Here $u_{\rm c}$ and $u_{\rm BH}$ denote
the centre and the black hole apparent horizon, respectively.}
\end{center}
\end{figure}

In the following, we focus on an ingoing null surface $v=v_{0}$,
as shown schematically in Fig.~\ref{fg:null}.
For a given matter stress-energy tensor, $a$
can be chosen arbitrarily on this surface and then $r$ is determined 
by Eq.~(\ref{eq:Ruu}). This implies that the form of the function $a$
on $v=v_{0}$ can be understood as a gauge freedom.

At the black hole apparent horizon the expansion 
along the outgoing null congruence vanishes, i.e. $r_{,v}=0$, while
at the cosmological apparent horizon the expansion 
along the ingoing null congruence vanishes, i.e. $r_{,u}=0$.
The cosmological apparent horizon is distinct from 
the cosmological particle horizon and defined in terms of 
the quasi-local properties of spacetime.
The radius of the cosmological apparent horizon is the Hubble
radius $H^{-1}$ in a flat Friedmann universe, as shown later.
From the definition (\ref{eq:misner-sharp}) of the Hawking mass, 
it is clear that $r=2m$ at these apparent 
horizons.
 
\begin{lm}
\label{lm:horizon}
The condition $2m/r<1$ holds if and only if $r_{,u}r_{,v}<0$.
The condition $2m/r=1$ holds if and only if $r_{,u}r_{,v}=0$.
The condition $2m/r>1$ holds if and only if $r_{,u}r_{,v}>0$.
\end{lm}
{\it Proof.} Trivial from Eq.~(\ref{eq:misner-sharp}). $\Box$
\begin{tm}
\label{tm:r_decreasing}
Suppose there exists $u_{0}$ such that $r>0$,
$r_{,v}>0$ and $r_{,u}<0$ at $(u_{0},v_{0})$.
Then if $T_{uu}\ge 0$, 
there exists a centre $u_{\rm c}$ ($>u_{0}$) 
such that $r(u_{\rm c},v_{0})=0$
and $r(u,v_{0})$ is a strictly 
decreasing function of $u$ 
for $u_{0}<u<u_{\rm c}$.
\end{tm}
{\it Proof.} 
Equation (\ref{eq:Ruu}) can be rewritten as
\[
\left(\frac{r_{,u}}{a^{2}}\right)_{,u}=-4\pi \frac{r}{a^{2}}T_{uu}.
\] 
The condition $r>0$ at $(u_{0},v_{0})$ implies that $r$ is non-zero
there, so
the left-hand side is negative on 
the null surface $v=v_{0}$.
Since $r>0$ and $r_{,u}<0$ at $(u,v_{0})=(u_{0},v_{0})$, we have 
$(r_{,u}/a^{2})(u,v_{0})\le (r_{,u}/a^{2})(u_{0},v_{0})<0$ for 
$u>u_{0}$ and $r>0$.
Since $a^{2}(u,v_{0})$ is subject to the gauge choice,
we can always assume that $a^{2}(u,v_{0})$ is bounded 
from below by a positive value. 
Thus $r_{,u}$ is bounded from above by a negative value.
Therefore, there exists $u_{\rm c}$ ($>u_{0}$) such that 
$r=0$ at $(u_{\rm c},v_{0})$ and such that $r(u,v_{0})$ 
is a strictly decreasing function of $u$ for $u_{0}<u<u_{\rm c}$.
$\Box$

\begin{lm}
\label{lm:bhah}
Suppose there exists $u_{0}$ such that $r>0$,
$r_{,v}>0$ and $r_{,u}<0$ at $(u_{0},v_{0})$.
If $T_{uu}\ge 0$ and 
there exist roots $\lambda$ such that $2m/r=1$ at $(\lambda,v_{0})$
for $u_{0}<\lambda<u_{\rm c}$, then $r_{,u}<0$ and $r_{,v}>0$ 
for $u_{0}<u<u_{\rm BH}$, where $u=u_{\rm BH}$ is the smallest root 
and corresponds to the 
black hole apparent horizon.
If there is no such root, then $r_{,v}>0$ and $r_{,u}<0$ 
for $u_{0}<u<u_{\rm c}$.
\end{lm}
{\it Proof.} Trivial from Lemma~\ref{lm:horizon} 
and Theorem~\ref{tm:r_decreasing}. $\Box$ 

\begin{tm}
\label{tm:m_decreasing}
Suppose there exists $u_{0}$ such that $r>0$,
$r_{,v}>0$ and $r_{,u}<0$ at $(u_{0},v_{0})$.
If $T_{uu}\ge 0$ and $T_{uv}\ge 0$,
then $m(u,v_{0})$ is a decreasing function of $u$
for $u_{0}<u<u_{\rm BH}$ if a black hole apparent horizon exists 
at $u=u_{\rm BH}>u_{0}$
and for $u_{0}<u<u_{\rm c}$ otherwise, where 
$r(u_{\rm c},v_{0})=0$.
$m(u,v_{0})$ is constant if and only if 
$T_{uu}=T_{uv}=0$.
\end{tm}
{\it Proof.} Since $r_{,u}<0$ and $r_{,v}>0$ 
in the relevant interval, 
equation~(\ref{eq:mu}) implies $m_{,u}\le 0$. $\Box$
 
\begin{tm}
\label{tm:upper_limit_bh}
Suppose that $T_{uu}\ge 0$ and $T_{uv}\ge 0$ and 
that there exists $u_{0}$ such that $r>0$,
$r_{,v}>0$ and $r_{,u}<0$ at $(u_{0},v_{0})$.
If a black hole apparent horizon exists at $u_{\rm BH}>u_{0}$,
then its area radius 
is less than or equal to $2 m(u_{0},v_{0})$.
Equality holds if and only if $T_{uu}=T_{uv}=0$.
\end{tm}
{\it Proof.}
From Lemma~\ref{lm:horizon} and Theorem~\ref{tm:m_decreasing},
we have $r(u_{\rm BH},v_{0})=2m(u_{\rm BH},v_{0})\le 2m(u_{0},v_{0})$.
$\Box$
\vspace*{1cm}

Here we comment on the energy conditions assumed above. 
For a perfect fluid, we have 
\begin{equation}
T^{ab}=\rho u^{a}u^{b}+p(g^{ab}+u^{a}u^{b}),
\end{equation}
where $u^{a}$, $\rho$ and $p$ are the four velocity, 
energy density and pressure, respectively.
The condition $T_{uu}\ge 0$ is equivalent to requiring $\rho\ge -p$.
The condition $T_{uv}\ge 0$ is equivalent to 
requiring $\rho \ge p$.
These two conditions together are
equivalent to the dominant energy condition $\rho\ge |p|$.
If we assume an equation of state of the form $p=k\rho$,
the dominant energy condition requires $-1\le k\le 1$ and $\rho\ge 0$.
For a massless scalar field $\Psi$, we have
 \begin{equation}
T^{ab}=\Psi^{,a}\Psi^{,b}-\frac{1}{2}g^{ab}\Psi^{,c}\Psi_{,c},
\end{equation}
and the conditions $T_{uu}\ge 0$ and $T_{uv}=0$ 
automatically hold.
Recall that a massless scalar field can be identified with a stiff fluid
(with the equation of state $p=\rho$)
if it is vorticity-free and has a timelike gradient~\cite{madsen1988}.
However, we do not need this equivalence in the following discussion.
 
\section{Application of upper limit to a PBH}
The above considerations are completely general, but we now focus on the situation in which one has a black hole 
in a cosmological background. To make the discussion precise, we
adopt a ``compensated'' PBH model, in which the 
black hole apparent horizon is contained within a perturbed region,
which is surrounded by an exact flat Friedmann 
solution beyond some (generally finite) radius. Various 
approaches have been used in this context.  One is to model the black hole 
as part of a closed Friedmann model, surrounded by a Schwarzschild void, so that the total mass at the edge of the 
void corresponds to that within a flat Friedmann model at the same
radius~\cite{nnp1978}. Another is to match the collapsing region 
to the flat Friedmann background by a shock-wave or a sound-wave, since this allows a discontinuity in the density 
or density gradient~\cite{ch1974}. If the PBH forms from density perturbations generated in a preceding inflationary 
phase, then the black hole could have a size 
larger than the particle horizon, but we do not consider such a situation in this section. 
 
We therefore have various possible situations, depending on the size of the matching radius. These possibilities 
correspond to the following definitions, in which the upper limit on the
matching radius is progressively reduced.
\begin{dn}
If the matching between the perturbed region 
containing the PBH and the flat Friedmann region is made at some finite 
radius, then we describe the PBH as ``locally produced''.
\end{dn}
If the perturbation results from a 
causal process, it is natural to assume that the 
perturbed region is inside the 
particle horizon of the surrounding flat Friedmann region. 
It should be noted that ``causal'', in this sense,  is distinct from ``physical'', since there exist physical 
mechanisms (such as inflation) 
for producing non-causal perturbations.
\begin{dn}
If the matching between the perturbed region 
containing the PBH and the flat Friedmann region is made inside the cosmological particle horizon of the flat 
Friedmann 
solution, then we describe the PBH as ``causally produced''.
\end{dn}
If the perturbation is produced by some hydrodynamical process,
it is natural to assume that the perturbation is confined 
within the cosmological sonic horizon, which is the distance traversed by a sound-wave since the big bang. In a flat 
Friedmann universe containing a perfect fluid with the equation of state 
$p=k\rho$
($0< k<1$),  this is just $\sqrt{k}$ times the particle horizon size. For a stiff equation of state $k=1$ or a 
massless scalar field, the cosmological sonic horizon 
coincides with the particle horizon. 
\begin{dn}
If the matching between the perturbed region 
containing the PBH and the flat Friedmann region is made inside the cosmological sonic horizon of the flat Friedmann 
solution, then we describe the PBH as ``hydrodynamically produced''.
\end{dn}
In fact, this is a very natural value for the matching radius, since the
existence of a sound-wave allows for a 
discontinuity  in the pressure and density gradients~\cite{ch1974}. 
Note that the 
cosmological sonic horizon is somewhat smaller than the Jeans length, which is the scale on which perturbations 
oscillate as sound-waves rather than grow.  As discussed in the Introduction, an overdense region needs to be bigger 
than the Jeans length to collapse against the pressure~\cite{carr1975} and this might seem to be incompatible with 
the last definition. However, in this argument, the relevant Jeans length is measured at the time when the overdense 
region has its maximum expansion and this is somewhat before the black hole itself forms.
 
We now calculate these scales explicitly for a flat Friedmann universe with 
\begin{equation}
ds^{2}=-dt^2+a^{2}(t)[d\chi^{2}+\chi^{2}(d\theta^{2}
+\sin^{2}\theta d\phi^{2})].
\end{equation}
In this case, the null coordinates are 
related to the conformal time and  
comoving radial coordinates $\eta$ and $\chi$ by
\begin{eqnarray}
u&=&\eta - \chi, \\
v&=&\eta + \chi,
\end{eqnarray}
where $a d\eta =dt$.
It is most convenient to derive expressions in terms of $u$ and
$v$ since this is more suitable for the application of the results 
obtained in the previous section.
In such coordinates the flat Friedmann 
spacetime for a perfect fluid with 
$p=k\rho$ is given by 
\begin{eqnarray}
a&=&C\eta^{2/(1+3k)}=  C\left(\frac{u+v}{2}\right)^{2/(1+3k)}, 
\label{eq:a_uv}\\
r&=&a\chi = C \left(\frac{u+v}{2}\right)^{2/(1+3k)}\frac{v-u}{2},
\label{eq:r_uv}
\end{eqnarray}
where $C(>0)$ is a constant of integration.
In the following we assume $0\le k\le 1$. 

\begin{figure}[htbp]
\begin{center}
\begin{tabular}{cc}
\subfigure[]{\includegraphics[scale=0.8]{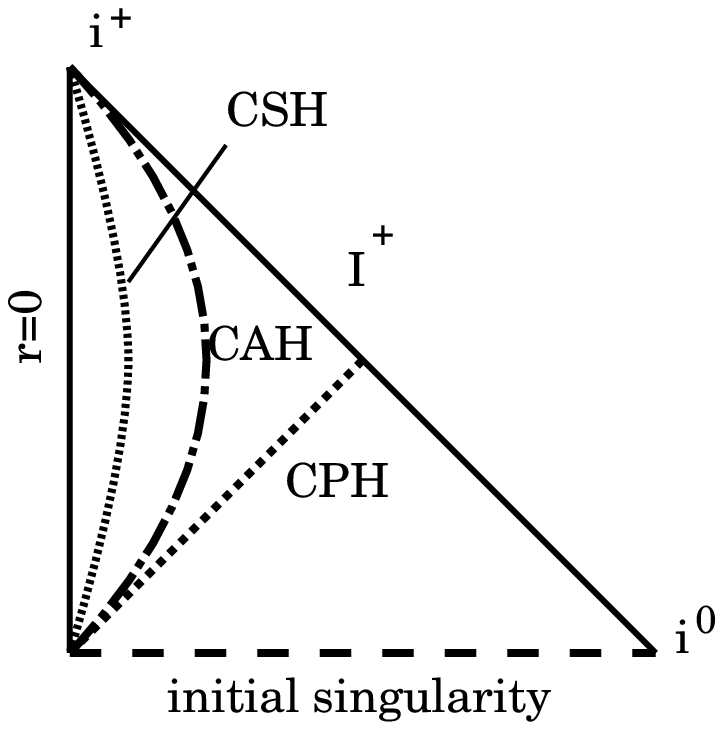}}
&\subfigure[]{\includegraphics[scale=0.8]{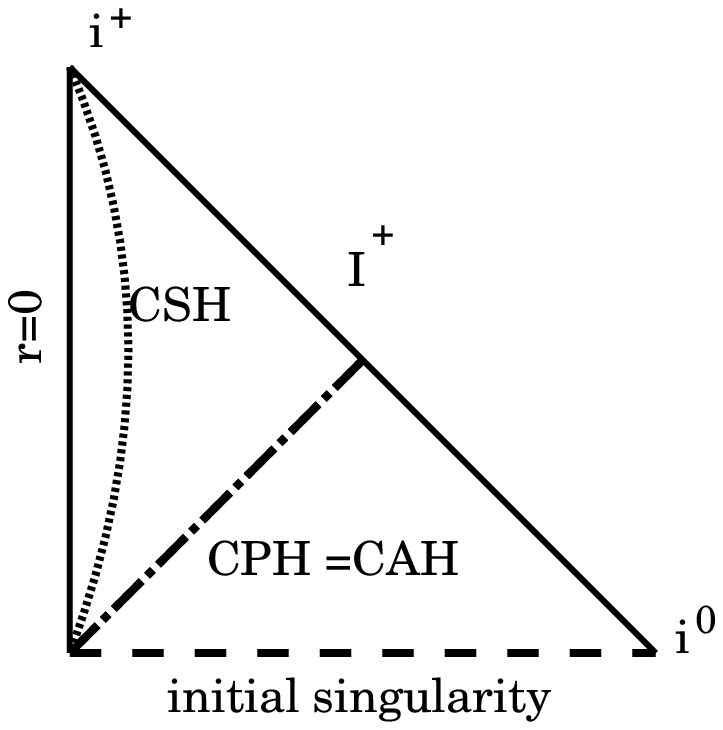}}\\
\subfigure[]{\includegraphics[scale=0.8]{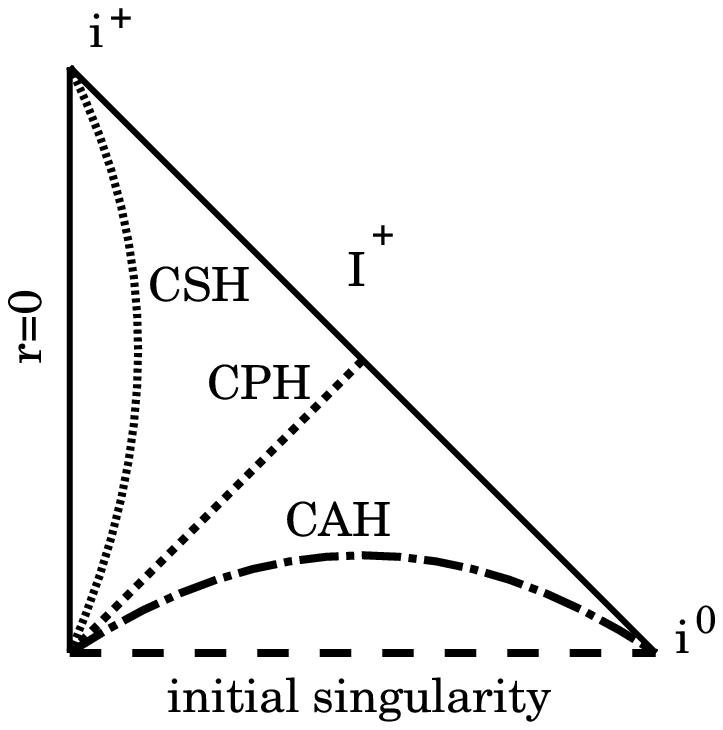}}
&\subfigure[]{\includegraphics[scale=0.8]{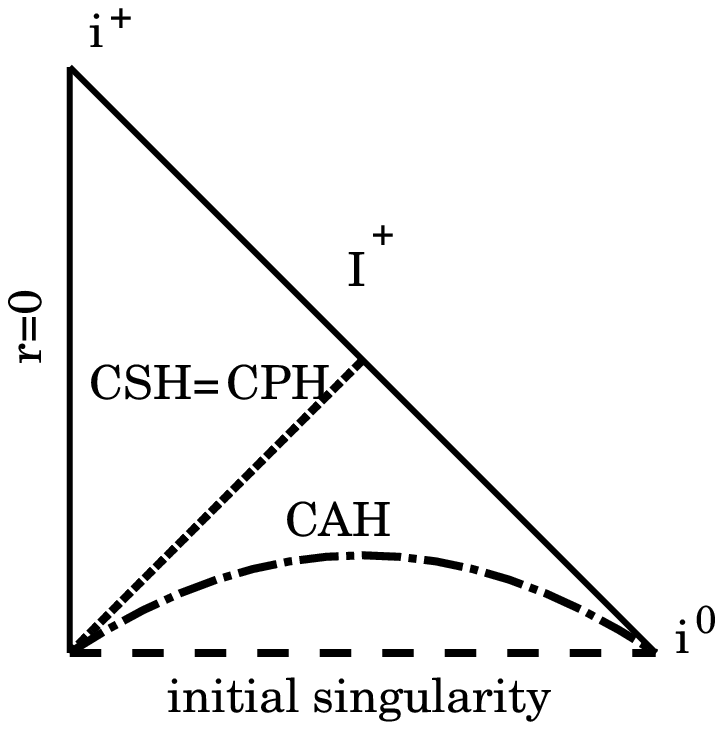}}
\end{tabular}
\caption{\label{fg:flat_friedmann}
The conformal diagrams for flat Friedmann 
spacetimes with a perfect fluid with the equation 
of state $p=k\rho$ for (a) $0<k<1/3$, (b) $k=1/3$,
(c) $1/3<k<1$ and (d) $k=1$ or a massless scalar field.
CAH, CPH, CSH denote the cosmological apparent horizon,
particle horizon and sonic horizon, respectively.}
\end{center}
\end{figure}

The cosmological particle horizon is given by
\begin{equation}
u_{\rm CPH}(v)=0,
\label{eq:u_cph}
\end{equation}
(i.e. $\eta = \chi$). 
Using Eqs.(\ref{eq:misner-sharp}), (\ref{eq:a_uv}) and (\ref{eq:r_uv}),
we obtain
\begin{equation}
\frac{2m}{r}=\left(\frac{2}{1+3k}\right)^{2}
\left(\frac{v-u}{v+u}\right)^{2},
\label{eq:2moverr}
\end{equation}
so the cosmological apparent horizon has the retarded time coordinate
\begin{equation}
u_{\rm CAH}(v)=-\left(\frac{k-1/3}{1+k}\right)v.
\end{equation}
Therefore, the cosmological apparent horizon is spacelike for $1/3<k\le 1$,
null for $k=1/3$ 
and timelike for $0< k <1/3$, as seen in Fig.~\ref{fg:flat_friedmann}.
For $0<k\le 1$, the cosmological 
sonic horizon has the retarded coordinate 
\begin{equation}
u_{\rm CSH}(v)=\left( \frac{\eta - \chi}{\eta +\chi}\right) v = \frac{1-\sqrt{k}}{1+\sqrt{k}}v,
\label{eq:u_csh}
\end{equation}
since the sound-speed is $\sqrt{dp/d\rho}=\sqrt{k}$. 
As seen in Fig.~\ref{fg:flat_friedmann}, 
the cosmological sonic horizon is timelike and 
inside the cosmological particle horizon for $0<k<1$,
while it is null and coincides with the cosmological particle 
horizon for $k=1$.
It is also null for a flat Friedmann spacetime with a massless scalar
field, so $u_{\rm CSH}=0$ in these cases.
A similar figure has been presented in~\cite{vws1999}. 

The area radii for these three cosmological horizons are written 
in terms of the advanced time coordinate $v$ as
\begin{eqnarray}
r_{\rm CPH}(v)&=&C\left(\frac{v}{2}\right)^{\frac{3(1+k)}{1+3k}}, 
\label{eq:r_cph_v}\\
r_{\rm CAH}(v)&=&C\left( \frac{1+3k}{2}\right) \left[\frac{2v}{3(1+k)}\right]
^{\frac{3(1+k)}{1+3k}}, 
\label{eq:r_cah_v}\\
r_{\rm CSH}(v)&=&C\sqrt{k}\left(\frac{v}{1+\sqrt{k}}
\right)^{\frac{3(1+k)}{1+3k}},
\label{eq:r_csh_v}
\end{eqnarray}
where 
we use the abbreviation $r_{\rm CAH}(v)=r(u_{\rm CAH}(v),v)$.
It is also convenient to express
these scales in terms of the Hubble parameter $H$.
Since this is defined as
\begin{equation}
H=\frac{1}{a^2}\frac{da}{d\eta}=\frac{2}{1+3k}\frac{1}{a \eta},
\end{equation}
we have
\begin{equation}
aH=\frac{1}{(1+3k)(u+v)}.
\end{equation} 
The area radii of the cosmological particle horizon,
apparent horizon and sonic horizon 
are then given by
\begin{eqnarray}
r_{\rm CPH}(v)&=&\frac{2}{1+3k}H^{-1}, \\
r_{\rm CAH}(v)&=&H^{-1}, \\
r_{\rm CSH}(v)&=&\frac{2\sqrt{k}}{1+3k}H^{-1}
\label{eq:r_csh_v_h}
\end{eqnarray} 
in a flat Friedmann universe.
Note that $H$ is here the Hubble parameter at the cosmological apparent 
horizon of the exact Friedmann background and this may not be the same 
as the Hubble parameter within the perturbed region itself.

\begin{figure}[htbp]
\begin{center}
\includegraphics[scale=0.8]{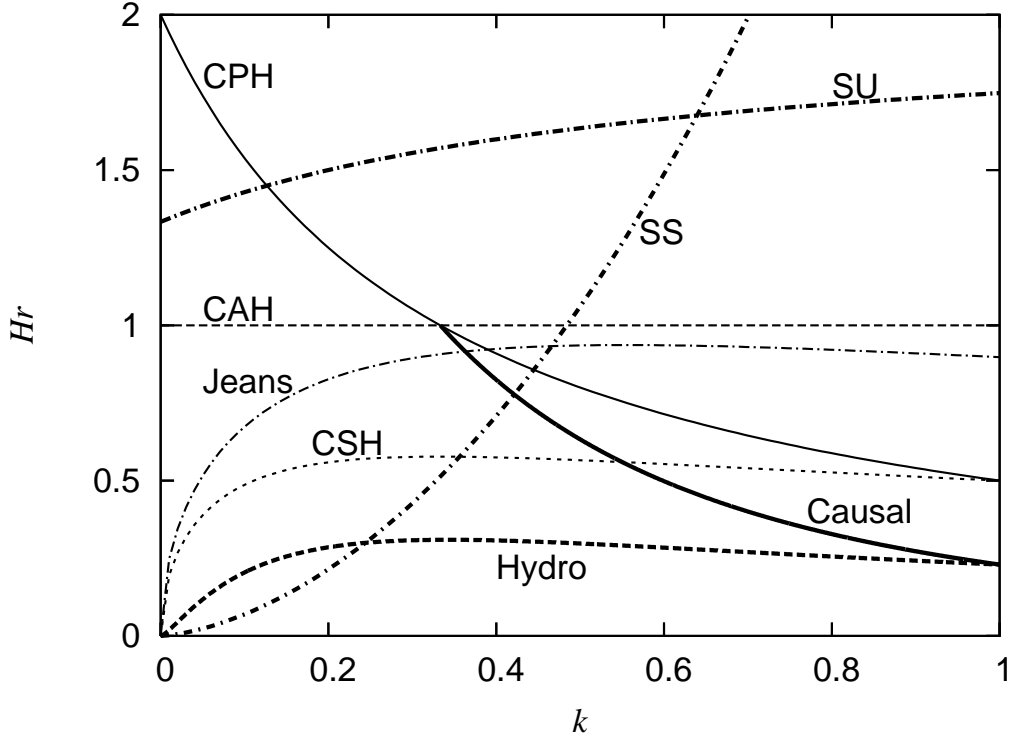}
\caption{\label{fg:horizons}
The characteristic physical scales in units of $H^{-1}$ 
for different values 
of $k$,
where $H$ is the Hubble parameter 
at the cosmological apparent horizon 
of the exact Friedmann background.
CPH, CAH and CSH 
denote the cosmological particle 
horizon, apparent horizon and sonic horizon, respectively.
``Jeans'' denotes the Jeans length
implied by Coles~\cite{coles1995}.
These four scales are plotted as thin lines.
Causal and Hydro denote the upper limits 
for causally and hydrodynamically produced PBHs, respectively.
SU denotes the upper limit which comes from 
the separate universe condition.
These three scales are plotted as thick lines.
SS denotes the black hole size
which the Newtonian argument of self-similar growth suggests.}
\end{center}
\end{figure}

Another important quantity is the Jeans scale, which is given by~\cite{ks1984}
\begin{equation}
\lambda_{\rm J}=\sqrt{\frac{\pi k}{\rho+p}},
\label{eq:jeans_ks1984}
\end{equation}
although this expression is strictly only valid for $\lambda_{\rm J}$
less than the cosmological apparent horizon size.
Below this scale density fluctuations oscillate as sound-waves rather than growing. However, it should be stressed that this is the Jeans length in the background universe, which is different from the Jeans length in an overdense region. It is usually assumed that a region can collapse to a black hole only if is larger than Jeans length at maximum expansion but, as discussed later, this is somewhat less than the value given by above. Since $\lambda_{\rm J}$ is a wavelength, the radius corresponding 
to Eq.~(\ref{eq:jeans_ks1984}) is
\begin{equation}
r_{\rm J}=\frac{1}{2}\lambda_{\rm J}=\pi\sqrt{\frac{2k}{3(1+k)}}H^{-1},
\end{equation}
where we have used the Friedmann relation 
\begin{equation}
H^{2}=\frac{8\pi}{3}\rho.
\end{equation}
However, we note that Coles~\cite{coles1995} gives a different equation for the Jeans length in a fluid with equation of state parameter $k$ and this leads to the expression
\begin{equation}
r_{\rm J}=\pi  \frac{4\sqrt {k}}{5+9k}H^{-1},
\label{eq:r_j_v_h}
\end{equation}
which we henceforth adopt.
This is always of order the sonic horizon but it is not exactly the
same. More precisely, the ratio of the physical scales given by
Eqs. (\ref{eq:r_csh_v_h}) and (\ref{eq:r_j_v_h})
is 
\begin{equation}
\frac{r_{\rm CSH}}{r_{\rm J}} = \frac{5+9k}{2\pi(1+3k)}.
\end{equation}
As $k$ goes from 0 to 1, this decreases monotonically
from 0.796 to 0.477.
We should note that the nature of the Jeans instability will be 
considerably modified for $k\agt 0.1$
because of the cosmological expansion~\cite{ks1984}.
These scales are depicted in Fig.~\ref{fg:horizons}.
It is interesting to note that $r_{\rm CPH}=r_{\rm CAH}$ 
at $k=1/3$ and $r_{\rm CSH}$ has a maximum there. 

At the cosmological particle horizon, equation~(\ref{eq:2moverr}) 
with $u=0$ implies
\begin{equation}
\frac{2m}{r}=\left(\frac{2}{1+3k}\right)^{2}.
\end{equation}
Therefore $2m/r< 1$ for $1/3< k\le 1$, $2m/r=1$ for $k=1/3$ 
and $2m/r>1$ for $0<k<1/3$.
This means that, for $1/3<k\le 1$,
the cosmological particle horizon is inside the cosmological
apparent horizon, i.e. $u_{\rm CPH}>u_{\rm CAH}$.
In this case, $r>0$, $r_{,u}<0$ and $r_{,v}>0$
at the cosmological particle horizon. Since a causally-produced PBH is
associated with a perturbed region inside the cosmological particle horizon,
Theorem~\ref{tm:upper_limit_bh} immediately implies the following:
\begin{tm}
\label{tm:upper_limit_pbh_fluid}
For a PBH causally produced from
a perfect fluid with $\rho\ge 0$ and $p=k\rho$ ($1/3<k\le 1$),
the ratio of the area radii of the
black hole and cosmological apparent horizons on
the same ingoing null surface is 
less than
\[
 \left(\frac{2}{1+3k}\right)^{3}
\left[\frac{3(1+k)}{4}\right]^{\frac{3(1+k)}{1+3k}}.
\]
\end{tm}
{\it Proof.} 
For $1/3<k\le 1$, we have $r>0$, $r_{,v}>0$ and $r_{,u}<0$ 
at $(u_{\rm CPH}(v),v)$.
Theorem~\ref{tm:upper_limit_bh} then gives
\begin{eqnarray*}
r_{\rm BH}(v)< 2m_{\rm CPH}(v) 
&=&\frac{2m_{\rm CPH}(v_0)}{r_{\rm CPH}(v)} 
\frac{r_{\rm CPH}(v)}{r_{\rm CAH}(v)}r_{\rm CAH}(v) \\
&=& 
\left(\frac{2}{1+3k}\right)^{2}\left( 
\frac{2}{1+3k}\right) \left[\frac{3(1+k)}{4}\right]^{\frac{3(1+k)}{1+3k}}
 r_{\rm CAH}(v), 
\end{eqnarray*}
where we have used Eqs.~(\ref{eq:r_cph_v}) and (\ref{eq:r_cah_v}). 
$\Box$ \\

\noindent
This upper limit is plotted in Fig.~\ref{fg:horizons}
and is always less than the cosmological particle horizon
for $k>1/3$.
Note that the ratio in Theorem~\ref{tm:upper_limit_pbh_fluid} 
peaks with a value of $1$ for a radiation equation of state 
and this is the most 
natural one for the early Universe.  
\begin{tm}
\label{tm:upper_limit_pbh_scalar}
For a PBH causally produced from
a massless scalar field, the ratio of the area radii of 
the black hole and cosmological apparent horizons
on the same ingoing null surface is 
less than $3\sqrt{6}/32$.
\end{tm}
{\it Proof.} This follows trivially by putting $k=1$ in 
Theorem~\ref{tm:upper_limit_pbh_fluid}.
$\Box$
\vspace*{1cm}
 
For $1/3<k\le 1$ or a massless scalar field,
once we have fixed the matching coordinate $u_{\rm match}$,
the radius of the black hole apparent horizon 
is maximized if there is a vacuum outside this horizon.
If we also change $u_{\rm match}$,
the upper limit on the black hole size is 
attained by the vacuum solution in the limit $u_{\rm match}\to u_{\rm CPH}$.
For $0<k\le 1/3$,
the radius of a causally produced PBH
could be as large as the cosmological particle horizon size. 
 
At the cosmological sonic horizon, we have 
\begin{equation}
\frac{2m}{r}=\left(\frac{2}{1+3k}\right)^{2}k
\end{equation}
and this is necessarily less than $1$
for $0<k\le 1$.
In this case, $r_{,u}<0$ and $r_{,v}>0$ at the cosmological sonic horizon.
For $k=1$, the cosmological sonic horizon coincides with 
the cosmological particle horizon.
Therefore, for a hydrodynamically-produced PBH,
we can prove the following.
\begin{tm}
\label{tm:upper_limit_pbh_fluid_hd}
For a PBH produced hydrodynamically from a perfect fluid with $\rho\ge 0$ and
$p=k\rho$ ($0<k\le 1$), the ratio of the area radii of 
the black hole and cosmological apparent horizons
on the same ingoing null surface is 
less than:
\begin{equation}
\left(\frac{2}{1+3k}\right)^{3}
\left[\frac{3(1+k)}{2(1+\sqrt{k})}\right]^{3(1+k)/(1+3k)}
k^{3/2}.
\label{eq:upper_limit_value}
\end{equation}
\end{tm}
{\it Proof.} 
For $0<k\le 1$, we have $r>0$, $r_{,v}>0$ and $r_{,u}<0$
at $(u_{\rm CSH}(v),v)$.
Then we have
\begin{eqnarray*}
r_{\rm BH}(v) < 2m_{\rm CSH}(v) 
&=&\frac{2m_{\rm CSH}(v)}{r_{\rm CSH}(v)}
\frac{r_{\rm CSH}(v)}{ r_{\rm CAH}(v)} r_{\rm CAH}(v) \\
&=& 
\left(\frac{2}{1+3k}\right)^{2} k 
\left( \frac{2\sqrt{k}}{1+3k}\right)
\left[\frac{3(1+k)}{2(1+\sqrt{k})}\right]^{3(1+k)/(1+3k)}
 r_{\rm CAH}(v),
\end{eqnarray*}
where we have used Eqs.~(\ref{eq:r_cah_v}) and (\ref{eq:r_csh_v}).
$\Box$\\ 

\noindent
This upper limit is also plotted in Fig.~\ref{fg:horizons}.
It is noted that the upper limit (\ref{eq:upper_limit_value})
takes a maximum value of 0.309
at $k\simeq 0.343$.

\section {Non-localised perturbations and separate universe condition}
We have considered PBHs formed from localised perturbations.
However, since the Friedmann universe will be perturbed everywhere 
in some situations (e.g. if the perturbations derive from
inflation), the above upper limit analysis 
may not always apply.
In this section we derive the condition that a PBH is part of 
our universe rather than being a separate closed universe.
This argument does not depend on whether the perturbation is 
localised or not.
For clarity we include factors $c$ and $G$ explicitly.
To make the discussion more precise, we assume that the overdense region 
which evolves to the PBH is part of a homogeneous closed Friedmann
model with
\begin{equation}
ds^{2}=-c^{2}dt^{2}+a^{2}(t)[d\chi^{2}+\sin^{2}\chi 
(d\theta^{2}+\sin^{2}\theta \phi^{2})].
\end{equation}
The cosmic scale factor within this region then evolves according to
\begin{equation}
\left(\frac{\dot{a}}{a}\right)^2 = \frac{8\pi G \rho}{3} 
- \frac{c^2}{a^2},
\label{eq:closed_Friedmann}
\end{equation}
where a dot denotes the derivative with respect to $t$.
This implies that the scale factor at maximum expansion is
\begin{equation}
a_{\rm max} = \left(\frac{3c^2}{8\pi G\rho}\right)^{1/2}.
\label{eq:a_max}
\end{equation}
For an equation of state parameter $k$, the background Friedmann density is
\begin{equation}
\rho_{\rm b} = \frac{1}{6\pi G (1+k)^2 t^2}.
\label{eq:background_density}
\end{equation}
If we denote the overdensity at maximum expansion by $\Delta \equiv (\rho_{\rm max}/\rho_{\rm b})$,  equation~(\ref{eq:a_max}) then gives
\begin{equation}
a_{\rm max} = \frac{3}{2}(1+k){\Delta^{-1/2}}ct = \Delta^{-1/2}cH^{-1}.
\label{eq:a_max_Delta_H}
\end{equation}
Since the maximum separation between two points on a spatial hypersphere
is $\pi$ times the radius of the hypersphere, the largest possible
radius for a region at maximum expansion is $r_{\rm max} =  \pi a_{\rm
max}$. 

The overdensity parameter $\Delta$ is expressed in terms
of $k$ as
\begin{equation}
\Delta=\frac{\pi}{4}\left[\frac{3(1+k)}{1+3k}\right]^{2}
\frac{\Gamma\left( \frac{3(1+k)}{2(1+3k)}\right)^{2}}
{\Gamma \left( \frac{2+3k}{1+3k}\right )^{2}},
\end{equation}
where the derivation is given in Appendix.
From Eq.~(\ref{eq:a_max_Delta_H}) our final expression for the separate
universe scale in units of $H^{-1}$ is
\begin{equation}
Hr_{\rm max}=\pi\Delta^{-1/2}= 2 \sqrt{\pi}\;  \frac{(1+3k)}{3(1+k)} \frac{
\Gamma \left( \frac{2+3k}{1+3k}\right )}{ \Gamma \left( \frac{3(1+k)}{2(1+3k)}\right)} .
\end{equation}
In a dust universe ($k=0$), this gives $4/3$. In a 
radiation universe ($k=1/3$), it gives $\pi/2$. In an (unphysical) 
universe with $k=\infty$, it gives $2$.
This expression is shown in Figure~\ref{fg:horizons}. 
Note that $r_{\rm min}$ exceeds the cosmological particle horizon for 
$k\agt 0.126$.
It should be stressed that, although we have presented our analysis in terms of the evolution of the overdense region, such a region does not evolve into separate universe. If it is a separate universe at maximum expansion, it is always separate.

We have noted that the Jeans length within an overdense region at maximum
expansion is less than the background Jeans length. Since the Jeans length
depends upon the inverse square root of the density, one might expect
the reduction to be of order $\Delta^{1/2}$, although this is not an exact
result since the expression given by Eq.~(\ref{eq:jeans_ks1984}) 
only applies in the linear
regime. The important point is that the usual lower limit on the size of a
black hole at formation is likely to be below the upper limits shown in
Fig.~\ref{fg:horizons}. 
Furthermore the existence of critical phenomena may circumvent the
Jeans limit anyway, since this permits black holes of arbitrarily small
mass.
\section{PBH accretion}
If we neglect the background expansion 
and assume a quasi-stationary flow,
a simple Newtonian treatment~\cite{zn1967} for a general fluid 
suggests that the accretion rate for a black hole of mass $M$ should be
given approximately by
\begin{equation}
\frac{dM}{dt} = 4\pi \alpha \rho R^2_{\rm A} v_{\rm s},
\label{eq:mdot}
\end{equation}
where $R_{\rm A}= G M/ v_{\rm s}^2$ is the accretion radius,
$v_{\rm s}$ is the sound-speed in 
the accreted fluid,
$\alpha$ is a factor of order unity which depends upon other 
physical features of the flow,
and factors $G$ and $c$ are now written explicitly. 
For a flat Friedmann universe containing fluid with equation of state $p=k\rho$, we have
\begin{eqnarray}
v_{\rm s}&=&\sqrt{k}c, \\ 
R_{\rm A} &=& \frac{GM}{kc^2}, \\  
\rho &=& \frac{1}{6\pi G(1+k)^2t^2}.
\end{eqnarray}
Therefore the accretion rate is
\begin{equation}
\frac{d M}{ d t} = \frac{2\alpha GM^2}{3k^{3/2}(1+k)^2 c^3 t^2}.
\end{equation}
This can be integrated to give
\begin{eqnarray}
\label{mass}
M &=& \frac{At}{1 + \frac{t}{t_{\rm f}} \left(\frac{At_{\rm f}}{M_{\rm f}} - 1 \right)},\\
A &=& \frac{3k^{3/2}(1+k)^2c^3}{2\alpha G},
\end{eqnarray}
where $M_{\rm f}$ is the black hole mass at the time $t_{\rm f}$ of
formation. If we 
assume that $M_{\rm f} = \xi A t_{\rm f}$ with $\xi <1$, 
then Eq.~(\ref{mass}) implies
\begin{equation}
M\rightarrow M_{\rm f}(1 - \xi)^{-1} \quad \mathrm{as} \quad t\rightarrow \infty.
\end{equation}
If $\xi \ll 1$, the black hole could not grow very much. However, if $\xi$ is 
close to 1, 
then the black hole 
could grow significantly. 
In other words, if the black hole mass is initially very close 
to $At_{\rm f}$, which is approximately the same as the mass
contained within the cosmological sonic horizon,
the black hole can grow considerably.
In particular, in the limit $\xi =1$, Eq.~(\ref{mass}) implies $M \propto t$, 
so the black hole grows at the same rate as 
the universe. 
The physical scale corresponding to the mass for which this analysis predicts self-similar growth is given by
\begin{equation}
r_{\rm f}H = 2\alpha^{-1}(1+k)k^{3/2}.
\end{equation}
This scale is also indicated in Fig.~\ref{fg:horizons} (for $\alpha =1$)
and one sees that in this case it violates the separate universe
condition for sufficiently large values of $k$.
This is another reason for regarding the Newtonian
prediction as suspect.

Since the above calculation neglects the effects of 
the cosmological expansion, 
one needs a relativistic calculation to check this. The Newtonian result 
suggests that one should look for a spherically symmetric {\it self-similar} 
solution, in which every dimensionless variable is a function of $z=r/t$, so 
that it is unchanged by the transformation 
$t \rightarrow at,\ r \rightarrow ar$ for any constant $a$. This problem has an 
interesting but rather convolved history. By looking for a black hole solution 
attached to an exact Friedmann solution via a sonic point, Carr \& Hawking 
first showed that there is no such solution for a radiation fluid~\cite{ch1974} and 
the argument was later extended~\cite{bh1978a} to a general $p= k \rho$ fluid with 
$0<k <1$. Lin et al.~\cite{lcf1976} subsequently claimed that 
there is such a solution in the special case $k =1$. However, Bicknell \& 
Henriksen~\cite{bh1978b} then showed that this solution is unphysical, in that the 
density gradient diverges at the event horizon. The solution can be completed 
only by attaching a Vaidya ingoing radiation solution interior to some surface 
(i.e. the scalar field has to turn into a null fluid). This is 
rather contrived, which
suggests that the black hole must soon become much smaller 
than the particle horizon, after which Eq.~(\ref{eq:mdot}) 
implies there will be 
very little further accretion. 
 
The present results are qualitatively consistent with
numerical calculations~\cite{nnp1978,np1980,nj1998,ss1999}. 
In particular, simulations~\cite{np1980} suggest that, 
the harder the equation of state, the more difficult it will be 
for a PBH to form and the smaller the mass accretion rate will be.
This may be understood in the context of the present analysis.
Even if the initial perturbation is larger than the cosmological 
apparent horizon, 
as may arise in the inflationary scenario,
it soon becomes comparable with the cosmological 
particle horizon, since the propagation speed of the perturbation 
is the sound speed, and this is slower than the light speed if $k<1$.
This means that the constraint on causally produced PBHs obtained 
also applies in this case. We conclude that the maximum mass and 
radius of these PBHs will decrease as the equation of state
becomes harder if $k>1/3$.

\section{Conclusion}
We have proved that a causally or hydrodynamically produced PBH 
cannot be as large as the cosmological 
apparent horizon for either a perfect fluid 
with a hard equation of state or a massless scalar field. 
It is clear that the result can be 
extended to a wider class of matter fields or 
to more general cosmological solutions.
We have also derived an exact expression for the 
maximum size that an overdense region can have without 
being a separate universe.
This is always larger than the cosmological apparent horizon and 
larger than the particle horizon size for 
a sufficiently hard equation of state.
Finally, we have considered accretion by PBHs but 
numerical simulations will be necessary to determine this precisely.
For the perfect fluid case, such simulations indicate a  small accretion rate.
For a massless scalar field, work is in progress 
by the present authors~\cite{hc2004}. 

\appendix*
\section{Calculation of the overdensity $\Delta$}
We here calculate the parameter $\Delta$ appearing in 
Eq.~(\ref{eq:a_max_Delta_H}). In the dust case this is well known to be $(3\pi/4)^2$ but we need to generalize this result.
For general $k$, we have $\rho \propto a^{-3(1+k)}$ and so 
Eq.~(\ref{eq:closed_Friedmann}) can be written as
\begin{equation}
\dot{a}^2 = A a^{-(1+3k)} - B
\label{eq:a_dot}
\end{equation}
where
\begin{equation}
A \equiv \dot{a_{\rm o}}^2\Omega_{\rm o} a_{\rm o}^{1+3k}, \;\;\;  B \equiv
 \dot{a}_{\rm o}^2(\Omega_{\rm o}-1) = c^2.
\label{eq:A_B}
\end{equation}
Here the subscript ``$_{\rm o}$'' indicates some initial epoch at which the overdensity is small, so that the density parameter $\Omega_{\rm o}$ is close to unity. 

By defining a new scale factor $b$ and a new time coordinate $\tau$ such that
\begin{equation}
b =  a^{1+3k}, \;\;\; d\tau = (1+3k) b^{\frac{3k}{1+3k}} dt, 
\label{eq:b_tau}
\end{equation}
we can transform Eq.~(\ref{eq:a_dot}) into the dust form:
\begin{equation}
\left(\frac{db}{d\tau}\right)^2 = \frac{A}{b} - c^2.
\label{eq:dbdtau}
\end{equation}
This has the well-known parametric solution:
\begin{equation}
b = \frac{1}{2} b_{\rm max}(1-\cos \eta), \;\;\; \tau =\frac{1}{\pi}
 \tau_{\rm max} (\eta - \sin \eta)
\label{eq:b_eta_tau_eta}
\end{equation}
where $\eta = \pi$ corresponds to the epoch of maximum expansion. 
Equations~(\ref{eq:b_tau}) and (\ref{eq:dbdtau}) then imply
\begin{eqnarray}
t &=& \frac{1}{1+3k} \int^{\tau}_{0}b^{-\frac{3k}{1+3k}}d\tau \nonumber \\
&=& \frac{1}{1+3k} 
\left(\frac{b_{\rm max}}{2}\right)^{-\frac{3k}{1+3k}}
\left(\frac{\tau_{\rm max}}{\pi}\right)
\int^{\eta}_{0} (1-\cos \eta)^{\frac{1}{1+3k}} d\eta .
\label{eq:t_eta}
\end{eqnarray}
Equations~(\ref{eq:a_dot}) and (\ref{eq:A_B}) give
\begin{equation}
\frac{a_{\rm max}}{a_{\rm o}} = \left(\frac{\Omega_{\rm o}}{\Omega_{\rm o} -1}\right) ^{\frac{1}{1+3k}},\;\;\;
\frac{b_{\rm max}}{b_{\rm o}} =\left( \frac{\Omega_{\rm o}}{\Omega_{\rm o} -1}\right) ,
\end{equation}
so we have
\begin{equation}
\frac{\rho_{\rm max}}{\rho_{\rm o}} = \left(\frac{a_{\rm max}}{a_{\rm o}}\right)^{-3(1+k)} =  \left(\frac{\Omega_{\rm o} -1}{\Omega_{\rm o}}\right) ^{\frac{3(1+k)}{1+3k}} .
\end{equation}
Since Eq.~(\ref{eq:background_density}) implies that the background density is given by
\begin{equation}
\frac{\rho_{\rm b}}{\rho_{\rm o}} = 
\left(\frac{t_{\rm o}}{t_{\rm max}}\right)^2,
\end{equation}
where $t_{\rm o}$ and $t_{\rm max}$ are given by Eq.~(\ref{eq:t_eta}), the overdensity can be expressed as
\begin{equation}
\Delta = \left(\frac{\Omega_{\rm o} -1}{\Omega_{\rm o}}\right)
 ^{\frac{3(1+k)}{1+3k}} 
\left[ \frac{\int^{\pi}_{0} (1-\cos \eta)^{\frac{1}{1+3k}}
 d\eta}{\int^{\eta_{\rm o}}_{0} (1-\cos  \eta)^{\frac{1}{1+3k}}
 d\eta}\right]^2 .
\label{eq:Delta_Omega_eta}
\end{equation}

Equation~(\ref{eq:b_eta_tau_eta}) implies that the integral limit $\eta_{\rm o}$, corresponding to the time $t_{\rm o}$, is given by
\begin{equation}
\eta_{\rm o} = \cos^{-1}\left(\frac{2-\Omega_{\rm o}}{\Omega_{\rm o}}\right) \approx
 2\sqrt {\frac{\Omega_{\rm o} -1}{\Omega_{\rm o}}}.
\label{eq:eta_Omega}
\end{equation}
The top integral in Eq.~(\ref{eq:Delta_Omega_eta}) can be expressed as
\begin{equation}
\int^{\pi}_{0} (1-\cos \eta)^{\frac{1}{1+3k}} d\eta=
2^{\frac{1}{1+3k}}\sqrt{\pi}\; 
\frac{\Gamma \left( \frac{3(1+k)}{2(1+3k)}\right)}{ 
\Gamma \left( \frac{2+3k}{1+3k}\right )} ,
\end{equation}
where $\Gamma(z)$ denotes the gamma function.
By using the fact that $\eta_{\rm o}$ is small, the lower integral 
in Eq.~(\ref{eq:Delta_Omega_eta}) can be simplified to
\begin{equation}
\int^{\eta_{\rm o}}_{0}\left(\frac{\eta^2}{2}\right) ^{\frac{1}{1+3k}} d\eta 
= 2^{\frac{2+3k}{1+3k}}\left(\frac{1+3k}{3(1+k)}\right) \left(\frac{\Omega_{\rm o} -1}{\Omega_{\rm o}}\right) ^{\frac{3(1+k)}{2(1+3k)}},
\end{equation}
where we have used Eq.~(\ref{eq:eta_Omega}). 
One observes that the terms involving
$\Omega_{\rm o}$ in the expression for $\Delta$ necessarily cancel to give
\begin{equation}
\Delta=\frac{\pi}{4}\left[\frac{3(1+k)}{1+3k}\right]^{2}
\frac{\Gamma\left( \frac{3(1+k)}{2(1+3k)}\right)^{2}}
{\Gamma \left( \frac{2+3k}{1+3k}\right )^{2}}.
\end{equation}

\acknowledgments
TH was supported from the JSPS.

\newpage 

\end{document}